\documentclass[dvips]{article} 
\usepackage{spie}   
\usepackage{psfig}   
\def\REF{}

\title{Topic Maps as a Virtual Observatory tool} 

\author{Ashish Mahabal\supit{a}, S. G. Djorgovski\supit{a},
Robert Brunner\supit{a} and Roy Williams\supit{a} 
\skiplinehalf 
\supit{a}Caltech, 1200 E California Blvd., Pasadena, CA, 91125, USA
}

\authorinfo{
\\
To appear in SPIE Annual Meeting 2001 proceedings (Astronomical Data Analysis)\\
Send correspondence to Ashish Mahabal.  E-mail: aam@astro.caltech.edu}

\begin{document} 
  \maketitle 

\begin{abstract}


One major component of the VO will be catalogs measuring gigabytes
and terrabytes if not more. Some mechanism like XML will be used
for structuring the information. However, such mechanisms are not
good for information retrieval on their own. For retrieval we use
queries. Topic Maps that have started becoming popular recently
are excellent for segregating information that results from a query.

A Topic Map is a structured network of hyperlinks above an information
pool. Different Topic Maps can form different layers above the same 
information pool and provide us with different views of it. This
facilitates in being able to ask exact questions, aiding us in looking
for gold needles in the proverbial haystack.

Here we will discuss the specifics of what Topic Maps are and how they
can be implemented within the VO framework.
\end{abstract}



\section{INTRODUCTION}
\label{sect:intro}  

The Astronomy and Astrophysics Survey Committee of the National Research Council
Decadal Report,
{\it Astronomy and Astrophysics in the new Millennium} [1]
has listed National Virtual
Observatory (NVO) as the highest priority item in the small program category.
The report reflects the community wide effort to harness the possibilities that
multi-terrabyte and multi-petabyte sky surveys are opening to the astronomy
community.
In parallel with the 
efforts here in the United States, various other places are involved in
contributing to a Virtual Observatory (VO). Here we will use the term
VO in a broad, generic sense.
Besides the traditional multi-wavelength surveys, the newer surveys
also encompass the time domain. The huge amount of data needs new tools,
techniques and a new way of doing astronomy.
In the next few pages we discuss Topic Maps (TM) as such a tool.

Different models and paradigms for a VO are under consideration. The commonality
between all these is that it will be web based and will provide transparent
access to a variety of datasets using tailored and/or user supplied tools. 
Many different issues have to be considered including various standards
for data storage, transmission and interchange.
Though different survey teams that will form part of the VO
will store the databases internally using 
different methodologies XML is the emerging standard to interface the data
with the rest of the world. This is mainly because the internet has become
the de facto medium for exchanging data.  It is for this reason that all
flavors of VO are web-based.
The most important feature of the VO will be to allow access to desired
information in a quick and systematic way.
XML allows for excellent structuring of information.
However XML itself is not good enough for
the intelligent maneuvering necessary
in the large dataspace. Topic Maps are an ideal
solution to this problem. With TMs one can define complex knowledge
structures and attribute them as metadata to information resources
allowing one to systematically organize knowledge on a variety of data subjects
such that the retrieval and sharing with other users is easy.

To use an analogy, TMs (plus a Topic Map Query language)
are to XML documents what SQL is to databases.
But TMs can do more than what SQL does for the following reasons:
(1) use of associations to qualify occurrences in TMs, (2) ability of TMs 
to describe abstract constructs, and
(3) the ability of TMs to exist independent of an information pool.
These three features together allow TMs to be employed for asking exact
questions and getting to the right datum in the data Himalayas.

Topic Maps are still new and are likely to develop a lot in the next few
years. They will be useful anyplace where Knowledge Organization (KO)
is useful. However, even at this early stage 
we can see three distinct areas in which Topic Maps can
serve the VO:
\begin{enumerate}
\item Enhanced searching for references and pointers to various
objects and keywords,
\item Searching for available observations of a given object, and
\item Providing knowledge organization in federated datasets.
\end{enumerate}

We begin by describing in Sect.~2 what Topic Maps are.
In Sect.~3 we outline a paradigm for the VO and describe how TMs become
the natural tool for various information extraction processes.
In Sect.~4 
we describe the three areas of Topic Map application mentioned above.
Finally, in Sect.~5,
we discuss the current status of topic
maps with respect to the availability of
resources and engines and compatibility issues.
In the appendix we present excerpts from an example Topic Map and the
Topic Map schema.

\section{What is a Topic Map}

A {\it Topic Map} is a collection of topics linked together by associations
between the topics. The topics can occur in different contexts and the 
associations qualify the occurrences. In XMLspeak
a Topic Map (which itself is an XML file)
is a structured network of hyperlinks above an information
pool (which is a set of one or more XML files).
Each node is a named topic. Associations between nodes are expressed
by links.  The named topics can be just about anything thereby
allowing a Topic Map to discuss abstract relationships between different topics.
Depending on the situation, different terms could be chosen to be
regarded as topics.

Here is a summary of various concepts associated with a topic:
\begin{itemize}
\item{
Each {\it Topic} has one or more {\it Topic Types}
e.g. topic NGC~4261 is of type ``galaxy''
and topic Europa is of type ``satellite''.
Topic types are also topics. Figure~1 shows a few topics in an information
pool.
}
\begin{figure}
\begin{center}
\begin{tabular}{c}
\psfig{figure=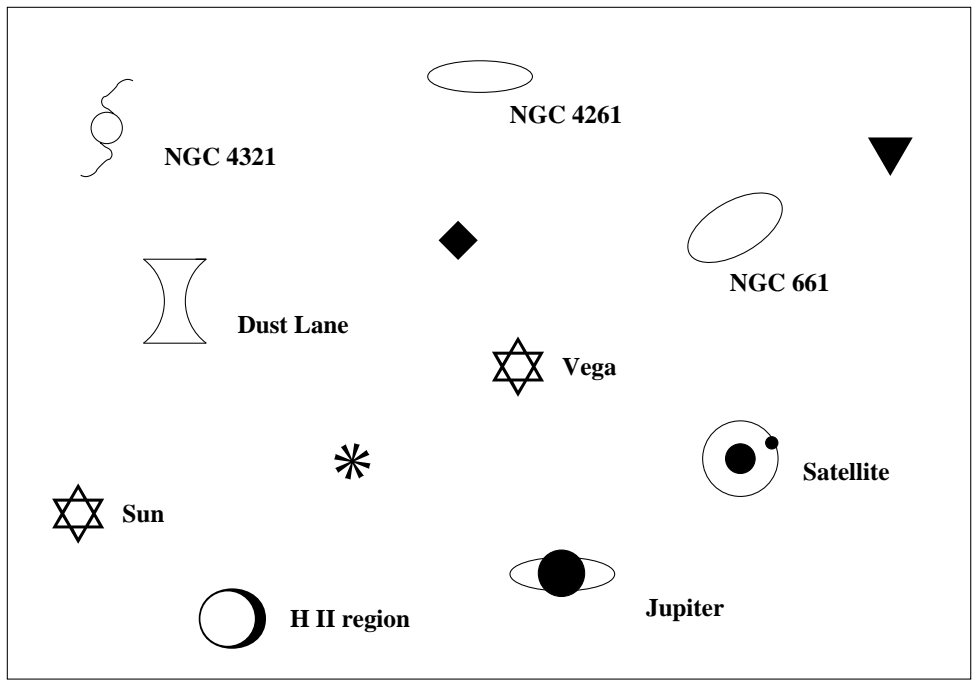,height=9cm}
\end{tabular}
\end{center}
\caption[figure 1] 
{ \label{fig:one}
A Topic Map can contain a large number of topics. A small selection
of astronomical topics is shown here which can go into forming a
Topic Map. The Topic Map itself is not shown. Different 
{\it topic types} have been shown using different symbols. A given
topic can belong to more than one topic type. Similarly, a topic
can have more than one name.
}
\end{figure} 

\item{
A topic also has one or more {\it Occurrences}. Each occurrence is a link
to a relevant information resource of the topic.
An occurrence is generally outside a Topic Map and is pointed to using some
addressing mechanism (e.g. XLink or XPointer).
Examples are: 
\begin{itemize}
\item  A paper on NGC~4261 entitled ``The blackhole of NGC~4261'' is an occurrence
for the topic NGC~4261,
\item  A mention that the dust lane in NGC~4261 is ``not similar to'' 
that of NGC~661 is another occurrence for NGC~4261,
\item  Presence of NGC~4261 in the NGC catalog is yet another occurrence,
\item  Mass of dust in NGC~4261 present in some compilation is yet another
possibility,
\item  A picture of NGC~4261 in the HST archive.
\end{itemize}

The occurrences lie in a different plane from the topics
as shown in Fig.~2.
}
\begin{figure}
\begin{center}
\begin{tabular}{c}
\psfig{figure=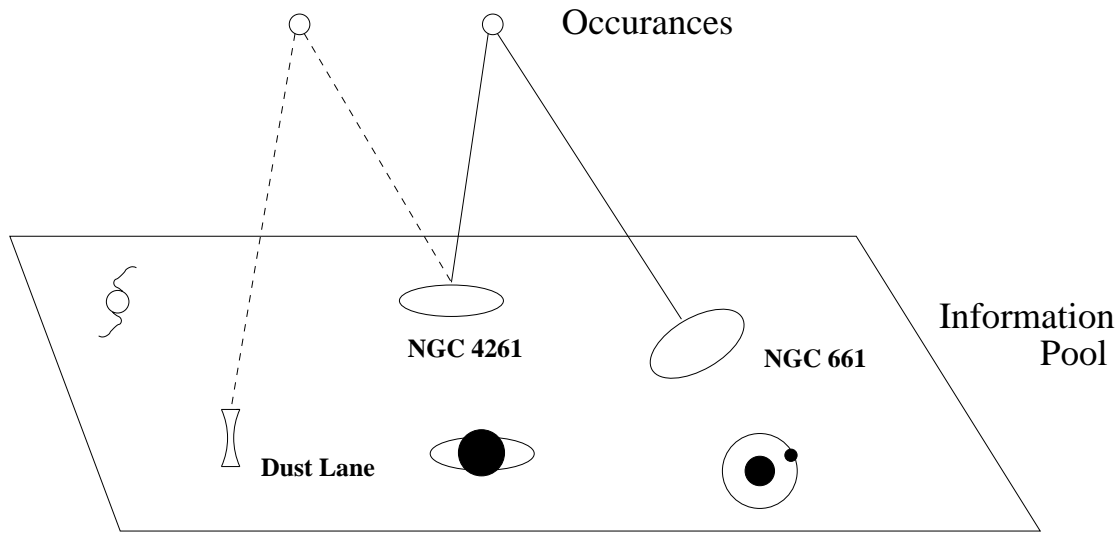,height=7cm}
\end{tabular}
\end{center}
\caption[figure 2] 
{ \label{fig:two}
Occurrences for a few topics from an information pool have been shown.
A single topic may have multiple occurrences in different contexts.
Here NGC~4261, an elliptical galaxy,
is shown to occur in connection with a dust lane and with
NGC~661, another elliptical galaxy.
}
\end{figure} 

\item{
Each topic occurrence has one or more {\it Occurrence Roles}:
These are topics in their own right e.g. consider 
``LMC is a satellite of the Galaxy''.
This is an occurrence for the topic LMC.
The role it plays is that of being a satellite.
Also, in the example above, different occurrences of NGC~4261 have
different occurrence roles viz. article, picture etc.
}

\item{
A topic also has an {\it Association} and an {\it Association Type}.
In the example of LMC above,
``is a satellite of'' describes the topic association between LMC and
our Galaxy. Thus associations denote the relationships between topics.
Associations can be grouped by their type. The grouping by association type
make large databases amenable to quick navigation.
In that respect Topic Maps are extensions of glossaries: the only relationship
that a glossary denotes is the definition. The associations between topics
are denoted using
link elements asserting the appropriate relationship. See Fig.~3 
for a few associations.

In Topic Maps, associations are by definition bilateral i.e. for each
association between two topics, a reverse association exists in the other
direction. LMC ``is a satellite of'' the Galaxy has the following
reciprocal association:
the Galaxy ``forms the main gravitational potential for'' or
``is orbited by'' the LMC.
\begin{figure}
\begin{center}
\begin{tabular}{c}
\psfig{figure=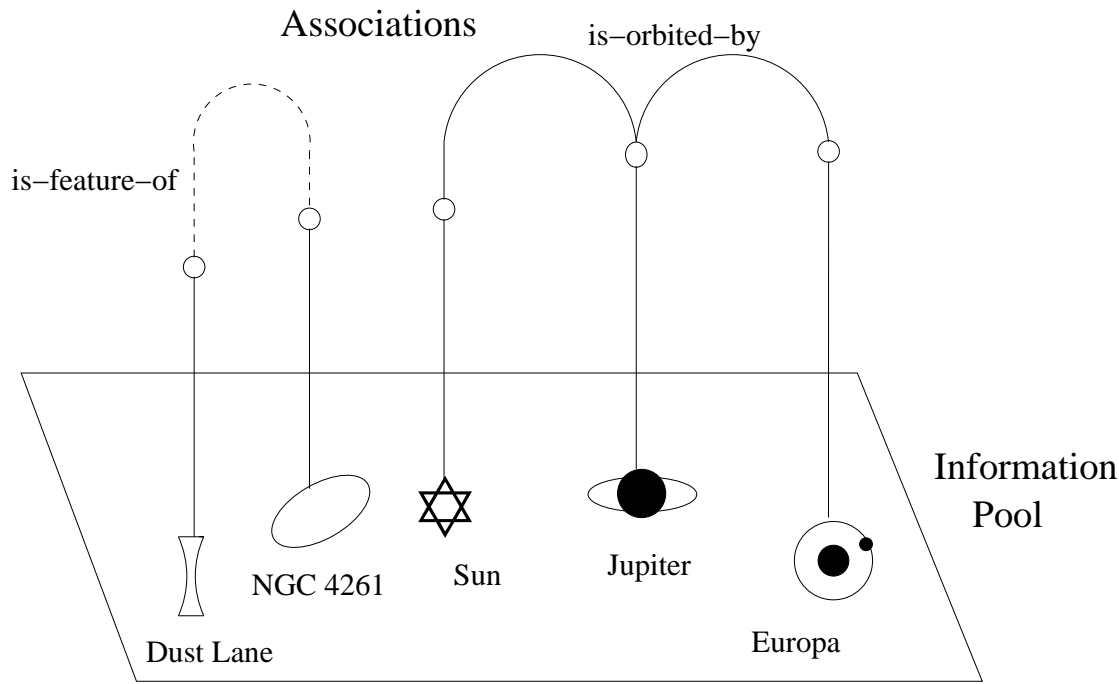,height=9cm}
\end{tabular}
\end{center}
\caption[figure 3] 
{ \label{fig:three}
Associations are a key feature of Topic Maps. Associations provide
links between different topics when a Topic Map is overlaid on an
information resource. In this figure two association types are shown:
(1) a dust lane ``is a feature of'' a galaxy, and (2) two associations
involving the ``is orbited by'' relationship.
}
\end{figure} 

Associations may or may not be symmetric $(a \rightarrow b => b \rightarrow a)$
e.g. Lyra $\zeta$ is the binary companion of Lyra $\iota$ is symmetric
with Lyra $\iota$ is the binary companion of Lyra $\zeta$, but
Europa is the satellite of Jupiter does not have a symmetric association
that is true.

Similarly, associations 
may or may not be transitive $(a \rightarrow b, b \rightarrow c
=> a \rightarrow c)$.
An example of a transitive association is that
Europa is part of the Jovian system
and the Jovian system is part of our Solar System together imply
that Europa is part of our Solar System.
An example of a non-transitive association is the following. Europa
is a satellite of Jupiter and Jupiter is a satellite of the Sun (in the broader
sense of the word satellite). However, Europa is not a satellite of the Sun.

To an extent, topic associations are like cross-references. However,
there is an important distinction: a cross-reference associates two items
within an information resource. On the other hand, an association links
two items independent of information resources. As a result, one can talk
of Topic Maps as independent information resources in their own right
(see Fig.~4).
A Topic Map can
then be overlaid on different information pools, or two Topic Maps
can be merged or interchanged.
}
\begin{figure}
\begin{center}
\begin{tabular}{c}
\psfig{figure=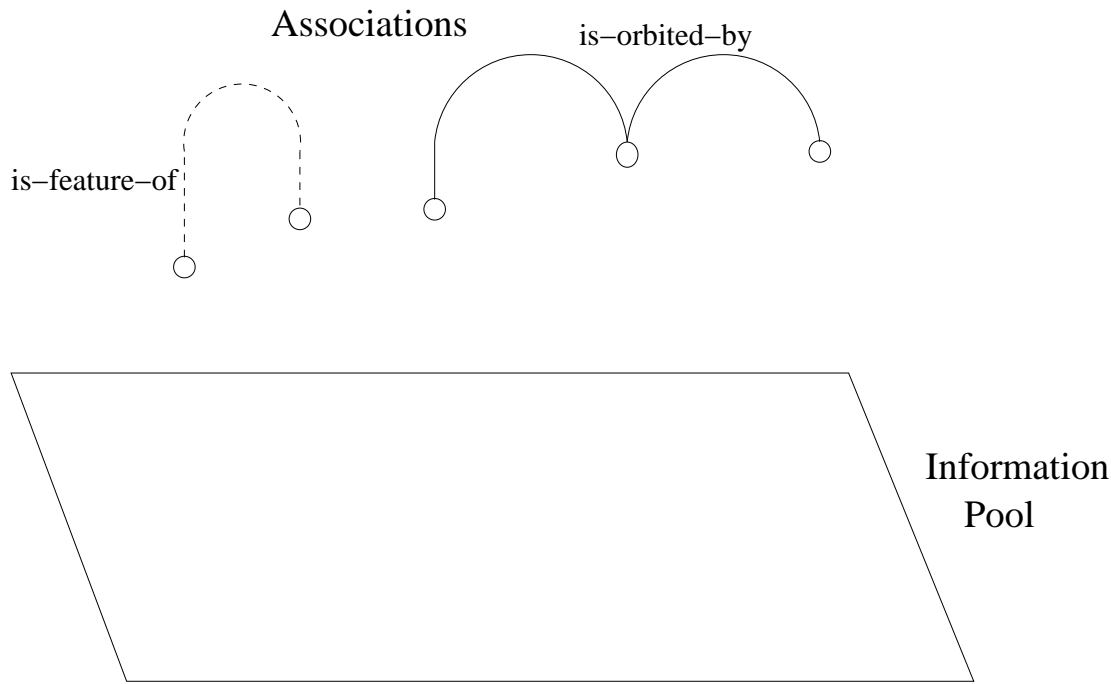,height=9cm}
\end{tabular}
\end{center}
\caption[figure 4] 
{ \label{fig:four}
The most important factor that distinguishes a topic association
from a cross-reference is its ability to exist despite the absence
of an information resource to go with it. Associations in a Topic Map
have a life on their own. In the figure the information pool is
blank. However, when a set of associations is defined as part of a
Topic Map, one can simply move it over another appropriate
information resource and make it work. This is what makes the merging
and interchange of Topic Maps easy.
}
\end{figure} 

\item{
A topic has a specific {\it Scope} in a given context.
It denotes the limit of validity of an assignment e.g. Europa is 
an astronomical body in its own right. We talk about it as a satellite only
with respect to its relationship with Jupiter.
Different themes define different scopes e.g. under the theme ``radio universe'',
NGC~4261 is a radio galaxy with FR II structure 
(where NGC~4261 is the public name associated with
3C270 -- see below) and under the theme
``optical universe'' NGC~4261 is an elliptical galaxy.
If there are two separate Topic Maps
for the radio universe and the optical universe,
they could be merged using the common or public identifiers like NGC~4261.
In astronomy the coordinate
positions available in different catalogs 
for the same object can vary slightly
and it will be more prudent to use RA-Dec pair for public names (for objects
having small proper motions in any case) and truncate the values to account
for the error bars in different catalogs to get associations.
}

\item{
A topic has one or more {\it Facets}. Facets are like properties and
help assign property-value pairs.
The difference with scoping is that it groups on properties of topics and
not on properties of the information resource.
Facets are the windows to the associated metadata and allow for
the filtering of information resources. Thus facets provide the metadata that
would have been provided by XML attributes. For example, for
an observation, this could
include attributes such as ``observer'', ``telescope'', ``instrument'',
``filter''. These provide additional classification possibilities
viz. {\it faceted} classification (though not in the same sense as used in
the KO jargon). One could then search for,
for example, not only
the observations of NGC~4261, but for {\it spectroscopic} observations of
NGC~4261 taken at a 4m+ telescope and so on.
}

\item{
Finally, a topic has a {\it Public Subject}. This can be used to combine
different topic names belonging to the same topic under different contexts.
Thus, NGC~4261 could be the public subject of 3C270. The public subject
is in addition to the {\it base name}, {\it display name}, or
{\it sort name} that a topic
can have. A point to note is that a topic need not have a name. That happens when the topic exists only as a reference.
}
\end{itemize}

Because of the different occurrence roles, main descriptions (important
pointers) of an object can be separated from the rest. When searching in a
large information pool this becomes a boon. Using different Topic Maps
we can denote different occurrences as more important in different contexts.
As a result, different Topic Maps provide us with different perspectives of the
same dataset.
In addition,
different TMs can be merged to provide a composite view of multiple
datasets. In some sense, TMs are to XML files what SQL is to traditional
databases. XML is good for structuring information, but it is not good for
retrieval on its own. TMs are good for segregating information that results
from a query. While tools have existed to deal with semantic networks and
knowledge structures, the presence of topic associations and occurrence
roles gives Topic Maps an extra edge in knowledge representation.

To a great extent TMs are like Resource Description Framework (RDF) which is likely to be the workhorse for Semantic Networks. However,
the main advantage of TMs is that they have stronger semantics than RDF
\REF (see [2] and [3]
for more details).
What that means is that a converter could, in principle, make an RDF
document from a Topic Map. However, the RDF interpreter will not be able
to make complete sense of the resulting document since the underlying
semantic structure will be lost.
The other factor that favors the use of 
TMs for the web based VO is that
while RDF can point to arbitrary resources, TMs are restricted to the web.

\section{A sketch of VO}

In this section we discuss an approximate common denominator paradigm
for the VO. The VO will be distributed and will allow access to a variety
of datasets. The tools available will not only enhance the capabilities
for doing science, but will also enable entirely new scientific
investigations. A block structure for the VO is depicted in Fig.~5.
The biggest block is shown to be the VO itself. It will comprise of
various tools that will allow users to play with the data. These can be simple
plotting tools or they can be complex cross-correlation tools. Another
component will be various interfaces that will allow users to interact with
the VO. There will be a multitude of codes available to do statistical
analysis, classification etc. These could be user supplied and optional.
Visualization portals will allow fly-throughs to look at multi-dimensional data
for searching,
for example, rare objects. Metadata translators will be at the heart of
all the above, perhaps hidden from view quietly converting various data
to the requested output formats. This is where XML will be hard at work
maintaining the structure of the data. All transactions will be through
some query mechanism. Many queries will result in very
large datasets. A tool is needed to make sense of such output data.
As mentioned before, Topic Maps fit the bill. 

The bulk of VO will be formed by
observational archives at different wavelengths and LSST-like surveys
involving the time domain. In addition, there will
be significant inputs from theory in form of models. Simulations
will contribute on an almost equal
footing in form of simulated datasets from which catalogs can be {\it derived}
and then compared with actual observations. A large part of the remaining input will be formed by follow-up
observations, stand alone observations by individual observers,
self-contained observations and deeper but less extensive surveys.

\begin{figure}
\begin{center}
\begin{tabular}{c}
\psfig{figure=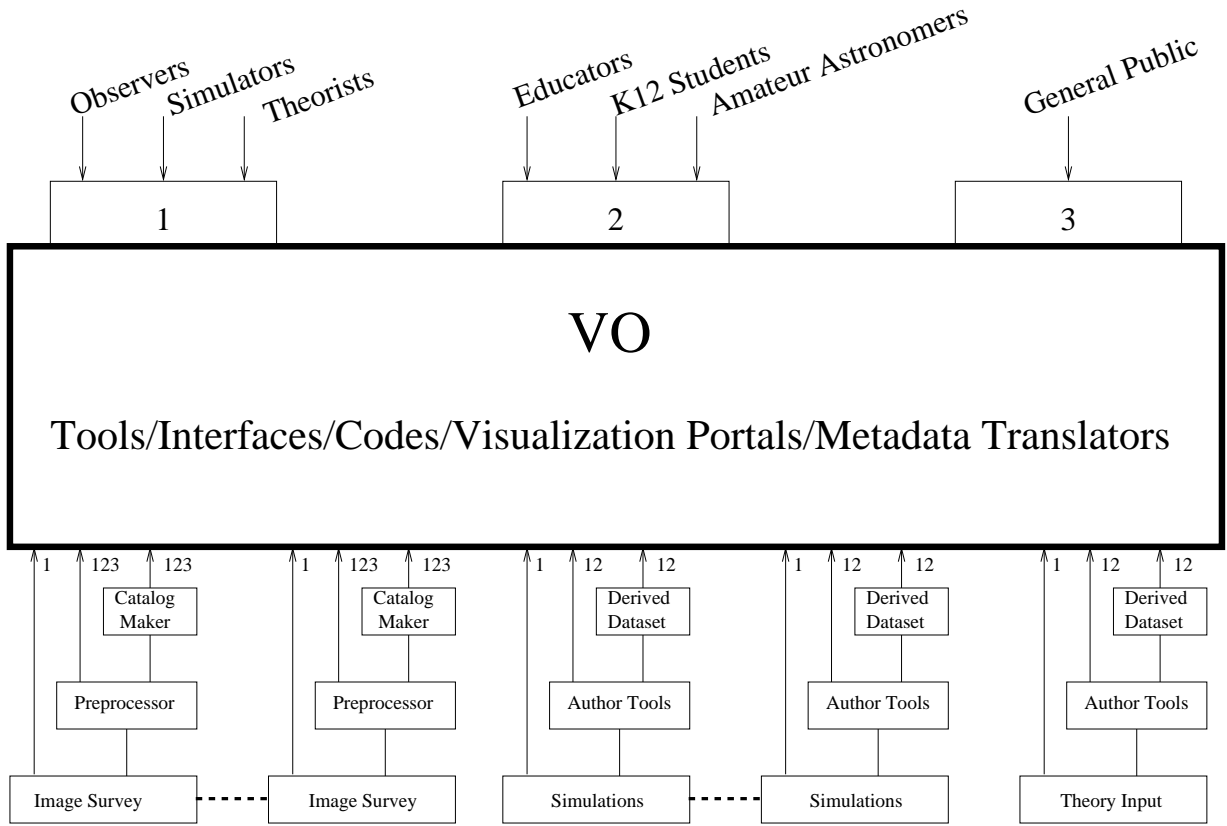,height=11cm}
\end{tabular}
\end{center}
\caption[figure 5] 
{ \label{fig:five}
A schematic of a Virtual Observatory. The main functionality of
the VO is to make data available from those who have it to those 
who will use it. The input data is in the form of observations, theory
and simulations. Many kinds of interfaces will allow the data to 
flow in different formats to researchers, educators and the general
public. Numbers 1, 2 and 3 denote the different access or clearance levels,
1 being the highest.
}
\end{figure} 

One can envisage the access to VO to be in 3 layers:
\begin{itemize}
\item
Clearance level 1 is highest and will open raw data to observers, simulators
and theorists. Since it can be misinterpreted by non-experts, and
since the preprocessed data will be separately available, the raw data
will not be open to non-researchers. The type of data available will include
observations, simulations, theoretically modeled.
\item
Level 2 shall be for educators, k12 students, amateur
astronomers etc.
This will provide access to processed data and catalogs and datasets/images
derived from simulations and theory. The general public will not have much
use for such data.
\item
Level 3 is for the general public and will provide the public outreach.
This is the most restricted category.
The general public 
can get catalogs, pretty pictures and processed data but no access to
simulations or to the raw data.
\end{itemize}

Topic Maps can play a vital role at both the input and output levels.
At the input, TMs can be used to merge different kinds of datasets
including those from observers and simulators.
Currently observers and simulators
do not speak identical languages in that their jargon differs (that is
also true for observers using different wavelengths like optical and
X-rays).
To merge their dataproducts,
we need to bring them on a common footing. Metadata can be used
to define public names (topics) of columns that mean the
same (using something like Column Descriptors used by CDS)
for the two communities and will provide an easy way to combine
the diverse databases into a derived product which could then be used 
on its own.

At the output, tailored TMs can be used to provide different views of
the data holdings of the VO, automagically implementing the different 
access levels for different type of users. The general public,
to quote an example, will not have to deal with data that they do not
understand, nor will educators have to deal with raw data.

Various different tools will exist to deal with frequent and large,
CPU consuming queries made by different types of users. Among the most
frequent queries are going to be queries like the following:
\begin{itemize}
\item
Has object {\it foo} been observed and if so, how do I get to its data,
\item
What information is available on topic {\it footoo}, and
\item
How do I correlate catalog X with catalog Y with such-and-such restrictions.
\end{itemize}
Topic maps provide an excellent mechanism to work in these areas.
We see how in the next section.

\section{Topic Maps for the Virtual Observatory}

In Sect.~2 we saw the different components of a Topic Map.
Here we sketch use scenarios for TMs using the common queries
mentioned in the previous section.
Excerpts of a corresponding Topic Map are in the Appendix.

\subsection{Observation Logs}
Consider a database of observational logs. It will be an important
component of the VO since it will tell us of available observations for any
object. The information pool here is formed by the logs of big surveys and those
of various observatories. For each observation obtained by 
individuals for different projects, some object will be the primary target.
This will become the main topic for such observations. Various
facets like telescope, observer, instrument and filter will get entered into
the XML file. Another important factor is the size of field covered (in
imaging mode, for instance).
Besides the obvious use of looking for objects that are the primary 
targets, Topic Maps, with the aid of appropriate catalogs, also assist 
the search for serendipitously observed objects.
We could thus use the field around a given object to search
for other objects with certain criteria as given in the example below.
When observations for an object are present (regular or serendipitous),
pointers to the location/source of the data will be returned. Each
observatory can maintain its own logs but so long as they are XML-compliant,
a single Topic Map will be able to work with all of them. Of course
for different instruments different Topic Maps could be used to decrease the
processing time. However, these different Topic Maps can then be merged
whenever needed.
All objects, observer names, instruments, telescopes etc. will form
separate topics. The associations between these will go into forming
the Topic Map. This will aid in asking exact questions like
``what spectroscopic observations of NGC~4261 have been done at
a 4m+ class telescope''.

Another example worth considering is of the following kind. Suppose one is 
looking for all cool stars for which spectra with the HST are available.
Currently, no straight forward mechanism is available to do this. When the two
databases are brought together, the natural tendency of the user is to
query out cool stars from an appropriate stellar database first and then
check which of those have HST spectra. A cheaper way would be to query
the HST database
for stellar spectra and then check which of those are cool stars.
With Topic Maps, this is how it can be done: A Topic Map on the stellar
database with stars as topics will have ``cool'' as a facet associated with the
cool stars. The HST database with spectra will similarly have the facet ``cool''
associated with spectra of cool stars. A merging of the two TMs (using
the names of stars for the merge parameter) and choosing those with
the facet ``cool'' will straightaway produce the desired listing.

An example of a more VO like 
search could be that of searching for QSO pairs. To do this
one could use a Topic Map (with QSOs as primary targets) over an information
pool consisting of observation logs. A merger of such a Topic Map with another
that has color outliers as its topics (with the catalogs from the corresponding
observations forming the information resource) could point to companion QSOs.


\subsection{Literature Database}

Originally Topic Maps were thought of in order to be able to merge different
indices, glossaries etc. Their use for merging different literature databases
to ease searches is analogous and hence immediately practical. The existing
databases like ADS, arXiv.org and those from individual journals maintain
a set of keywords and articles associated with such keywords. If Topic Maps
are used over such an information pool one can make use of occurrence roles
more effectively to get better results. When we search for ``dust lane galaxy''
in just the titles of papers, a lot of relevant articles are missed. If the
search is generalized to an occurrence in the abstract or text, far too many
and largely irrelevant answers are returned. In the later case, all
different occurrence roles and types are being returned in a pot pouri.
If only the keywords are stored as an XML file and a Topic Map with
``dust lane galaxy'' as a topic is overlaid, more appropriate results will
get returned. 

From the abstracts, keywords and classifications associated
with each document in a literature database,
a large number of topics can be chosen to form
a Topic Map. 
Additionally, each topic will have a number of facets associated with it.
Thus the Topic Map forms an additional layer
of structure over the already existing one. This allows one to maintain 
a view of the underlying knowledge without changing the
structure underneath. Different Topic Maps i.e. those with different
sets of topics and associations provide different views of the same 
underlying knowledge making it easier to search the structure and to get
to what one needs in a quick and precise way. In addition, browsing and
exploring the underlying knowledge structure becomes easier for the users.
However, for this to happen, proper Topic Maps need to be created with the
help of information specialists working closely with astronomers.

\subsection{Federating Datasets}
VO will not only make science easier to do, but it will in fact enable 
several different scientific investigations that would not be 
possible otherwise. One important component in the enabling nature
is the federation of large datasets. Though currently all the mechanism
is not in place, one of 
the grandest use of Topic Maps could be to provide knowledge organization
of the very large-sized federated datasets.
The access level to VO will determine the type of queries a user can run
on different databases. One of the most important queries will involve
cross-correlations between different datasets e.g. a sky-survey at
optical wavelengths like DPOSS and
another at radio wavelengths like FIRST. Federating two such datasets will
be the most basic step. It is easy if the two databases are of the same type.
An SQL query could access the necessary columns from the two and merge them
into a single derived table e.g. RA, Dec, $r_{mag}$ from DPOSS, and RA, Dec,
$S_{408}$ from FIRST.
However, different sky survey archives tend to have different databases: 
relational, objectivity etc. making it nontrivial to extract
information uniformly and merge it meaningfully. Topic Maps can provide a
mechanism that will make it possible. 
One could use the column names in the databases as topics
with all the synonyms as secondary names, and with wavelength, flux 
conversions included as associations (indeed, all kinds of conversions
can exist as associations e.g. ``is-same-as'', ``is-analogous-to'', ``is-identical-to'' etc.). However, as noted towards the end
of Sect.~5, Topic Maps have not really been tested for scalability to the 
extreme scales and one may have to tread cautiously.

Scalable or not, TMs will still work with huge databases when it is the
attributes rather than the objects that form the topics. Databases that
contain billions of objects will still contain at most hundreds of
attributes per object. A TM developed on the attributes will still be
important.

The web-face of the databases would again be XML. Topic Maps could utilize
the metadata associated with the information pool (database) to overlay
complex structure in a systematic way to organize the knowledge therein.
Again, different Topic Maps would provide different views and hence 
structuring of the knowledge without actually modifying the database.

\section{Current status, future prospects, and conclusions}

Though Topic Maps started life as a standard for software documentation 
in 1991 (HyTime Hypertext 1991 conference in San Antonio)
the progress has been slow. It is only when
XML matured that their real potential has been noticed. In 1996 Topic Maps
became a work item in ISO's SGML working group resulting in a Topic Map
standard published in 1999. Since then a number of commercial organizations
have taken an interest in Topic Maps and progress is expected to be brisk.
\REF More history can be found in [2].

A lot still needs to be done and hence this article is still more of an outline
than a report on a finished project:
\begin{itemize}
\item The tool support needed is immense. There will
have to be tools for Topic Map design, creation, merging, exchanging
and overlaying. Each of these processes are big goals in themselves e.g.
designing a consistent Topic Map needs careful definitions of different
types and roles of topics, scopes have to be defined and associations
and occurrences of different topics explicitly coded.
A visualization tool is necessary to extend a Topic Map and
to merge it with another one and even to check the different
queries that could be made with one.
Merging itself involves defining import/export methods and carefully
designed topic naming constraints.

\item ISO/IEC 13250 has declared interchange format for Topic Maps. However,
issues like transitivity and symmetry (see Sect.~2)
are not part of the standard
and browsers could presently include monopolized versions.

\item With multi-million node Topic Maps, powerful graphic interfaces will
be needed for navigation. There will have to be slots for user-definable
Topic Maps too.

\item Implementation issues like object model design and the storage and
searching of TMs need careful consideration.
\end{itemize}

Work is in progress on these fronts at different places. Currently
the best way to keep oneself updated on these issues is by following
the egroups on these topics, keeping an eye on the related web sites,
and perhaps actually getting involved.
It is important for astronomers to participate especially in making sure
that the functionality drivers deliver.

Some of the sites to look out for (and this is not an exhaustive list by
any means) are given below.
\begin{itemize}
\item http://www.topicmaps.org/xtm/1.0/ has the latest version of 
	TopicMaps.Org Specification (check also links on that page).
\item http://www.ontopia.net/. Ontopia has {\it tmproc}
	[4] 
	an implementation in Python.
\item http://www.mondeca.com/. Mondeca too has started supplying
	software components that are based on Topic Maps.
\item http://www.infoloom.com/. Infoloom and their tutorial at
	www.topicmaps.net. They also arrange conferences on
	Topic Maps (http://www.gca.org) and have a good FAQ
	(http://www.infoloom.com/tmfaq.htm).
\item http://k42.empolis.co.uk/tmv.html. {\it Topic Map View} from Empolis.
\item http://www.techquila.com/tm4j.html.
	{\it TM4J}: A Topic Map engine for Java.
\end{itemize}

Some questions that still need attention
(see [5]):
\begin{itemize}
\item Work on Knowledge Organization (KO) has been going on for
years. How best to translate those KO structures into meaningful
TM syntax without reinventing the wheel.
\item We still do not know how scalable Topic Maps are with respect to
layered semantic interoperability.
\end{itemize}

The work that we have currently undertaken is to build a set of Topic Maps
for the 100 galaxies closest to our Galaxy. The information pool is formed by 
the data from NED on these objects (in XML format devised by CDS) and
references from ADS. The work involves constructing Topic Maps from the 
viewpoint of an astrophysicist, astronomer, librarian, college student etc.
and is still at an elementary stage. As progress is made, it will be posted at:
http://www.astro.caltech.edu/$\sim\!$aam/science/topicmaps/.

To summarize, Topic Maps are emerging as a new and powerful tool
for knowledge organization in very large information resources.
The concept behind Topic Maps is that of a topic being used to access
information about it based on the underlying knowledge. The greatest 
advantage of Topic Maps is that they can be maintained independent of 
any information resources they are combined with, thus making them
information assets in their own right. Another advantage is that
different Topic Maps provide different views of the same information resource
and different Topic Maps can be merged and exchanged. This facilitates
the generation of an overall structure for multiple information resources
without changing the resources themselves. Various attempts are underway
to create tools necessary for Topic Map design, creation, maintenance
and merging.

\section*{APPENDIX}
\subsection*{Excerpts from an Example Topic Map}

\begin{verbatim}
!	Defining an object
<topic id="ngc4261">
        <instanceOf><topicRef XLink:href="#galaxy"/></instanceOf>
        <baseName>
                <baseNameString>NGC 4261</baseNameString>
        </baseName>
        <occurrence>
                <instanceOf><topicRef XLink:href="#image"/></instanceOf>
                <resourceRef XLink:href="http://ned/ngc4261.gif"/>
        </occurrence>
</topic>
!	Defining another object
<topic id="ngc4321"> 
        <instanceOf><topicRef XLink:href="#galaxy"/></instanceOf>
        <baseName>
                <baseNameString>NGC 4321</baseNameString>
        </baseName>
        <occurrence>
                <instanceOf><topicRef XLink:href="#image"/></instanceOf>
                <resourceRef XLink:href="http://ned/ngc4321.gif"/>
        </occurrence>
</topic>
!	Defining a feature
<topic id="dustlane">
        <instanceOf><topicRef XLink:href="#feature"/></instanceOf>
!        NO BASE NAME
        <occurrence>
                <instanceOf><topicRef XLink:href="#ngc4261"/><instanceOf>
                <resourceRef XLink:href="http://ned/ngc4261.ps"/>
        </occurrence>
</topic>
!	Defining another feature
<topic id="h2region">
        <instanceOf><topicRef XLink:href="#feature"/></instanceOf>
!        NO BASE NAME
        <occurrence>
                <instanceOf><topicRef XLink:href="#ngc4321"/><instanceOf>
                <resourceRef XLink:href="http://ned/ngc4321.ps"/>
        </occurrence>
</topic>

!	Defining an association
<association>
        <instanceOf><topicRef XLink:href="is-part-of"/></instanceOf>
        <member>
                <roleSpec><topicRef XLink:href="#object"/></roleSpec>
                <topicRef XLink:href="#ngc4261"/>
        </member>
        <member>
                <roleSpec><topicRef XLink:href="#feature"/></roleSpec>
                <topicRef XLink="#dustlane"/>
        </member>
</association>
!	Defining another association
<association>
        <instanceOf><topicRef XLink:href="is-part-of"/></instanceOf>
        <member>
                <roleSpec><topicRef XLink:href="#object"/></roleSpec>
                <topicRef XLink:href="#ngc4321"/>
        </member>
        <member>
                <roleSpec><topicRef XLink:href="#feature"/></roleSpec>
                <topicRef XLink="#h2region"/>
        </member>
</association>
\end{verbatim}

\subsection*{From the TM schema (ISO 13250)}

\begin{verbatim}
topicmap
        topic
                topname
                        basename
                        dispname
                        sortname
                occurs
        assoc
                assocrl
        facet
                fvalue
        addthms
\end{verbatim}
\acknowledgments     
This work was supported in part by
the NASA AISRP grant NAG5-9482.
Many thanks to Vidyullata Mahabal for help with the figures.


\section*{REFERENCES}
{[1] Astronomy and Astrophysics in the new Millennium, 2000, 
http://www.nap.edu/books/030970317/html/} \\
{[2] Pepper, S., Navigating haystacks and discovering needles, 1999, \\
http://www.ontopia.net/topicmaps/materials/mlangart.pdf} \\
{[3] Pepper, S., Topic Maps and RDF: A first cut, 2000, 
http://www.ontopia.net/topicmaps/materials/rdf.html} \\
{[4] tmproc: A Topic Map engine (in Python), 1999-, 
http://www.ontopia.net/software/tmproc/} \\
{[5] Sigel, A., Towards Knowledge Organization with Topic Maps, 2000, \\
http://www.gca.org/papers/xmleurope2000/papers/s22-02.html} \\

\end{document}